\documentclass[twocolumn,showpacs,preprintnumbers,amsmath,amssymb]{revtex4}

\newcommand{\new}[1]{ #1}

\usepackage{graphicx}
\usepackage{dcolumn}
\usepackage{bm}

\newcommand{\De}{\Delta}

\newcommand{\eps}{\epsilon}

\newcommand{\Om}{\Omega}

\newcommand{\p}{\partial}
 
\newcommand{\txt}{\textstyle}

\newcommand{\beq}{\begin{equation}}
\newcommand{\eeq}{\end{equation}}
\newcommand{\ba}{\begin{array}}
\newcommand{\ea}{\end{array}}
\newcommand{\bea}{\begin{eqnarray}}
\newcommand{\eea}{\end{eqnarray}}
\newcommand{\bi}{\begin{itemize}}  
\newcommand{\ei}{\end{itemize}}
\newcommand{\ben}{\begin{enumerate}} 
\newcommand{\een}{\end{enumerate}}

\newcommand{\half} {{\txt \frac{1}{2}}}
\newcommand{\third}{{\txt \frac{1}{3}}}

\newcommand{\twothirds}{{\txt \frac{2}{3}}}

\newcommand{\feyn}[1]{
  \setbox0=\hbox{\ensuremath{#1}}
  \hbox to\wd0{\hbox to0pt{\hbox to\wd0{\hss/\hss}\hss}\box0}}

\newcommand{\diag}{{\rm diag}}

\newcommand{\Qtilde} {{\tilde Q}}

\begin{document}

\preprint{MIT-CTP-3447}
 
\title{Gapless Color-Flavor-Locked Quark Matter}

\author{Mark Alford}%
\affiliation{Physics Department, Washington University,
St.~Louis, MO~63130, USA}
\author{Chris Kouvaris}\author{Krishna Rajagopal}
\affiliation{Center for Theoretical Physics, Massachusetts Institute of
Technology, Cambridge, MA~02139, USA}

\date{Nov 2003}

\begin{abstract}
In neutral cold quark matter that is sufficiently dense that the
strange quark mass $M_s$ is unimportant, all nine quarks (three
colors; three flavors) pair in a color-flavor locked (CFL) pattern,
and all fermionic quasiparticles have a
gap. We argue that as
a function of decreasing quark chemical potential $\mu$ or increasing
$M_s$, there is a 
quantum phase transition from the CFL 
phase to a new ``gapless CFL phase'' in which only seven quasiparticles
have a gap. The transition occurs
where $M_s^2/\mu\approx 2\Delta$, with $\Delta$ the gap
parameter.  Gapless CFL, like CFL,
leaves unbroken a linear combination $\tilde Q$ of electric and color
charges, but it is a $\tilde Q$-conductor with 
a nonzero electron density. These electrons and
the gapless quark quasiparticles make the low energy effective
theory of the gapless CFL phase and, consequently, its
astrophysical properties
qualitatively different from that of the CFL phase,
even though its $U(1)$ symmetries are the same.  Both gapless
quasiparticles have quadratic dispersion relations at the
quantum critical point. For values of $M_s^2/\mu$ above the quantum critical
point, one branch has conventional linear dispersion relations
while the other branch remains quadratic, up to 
tiny corrections.

\end{abstract}


\maketitle



We know a lot about the properties of cold quark matter
at sufficiently high baryon density from first principles.
Quarks near their Fermi surfaces pair, forming a color
superconductor~\cite{reviews}.
In this letter we study how the favored pairing pattern at
zero temperature
depends on the strange quark mass $M_s$, or equivalently
on the quark chemical potential $\mu$, using the 
pairing ansatz~\cite{Alford:1998mk}
\begin{equation}
\langle \psi^\alpha_a C\gamma_5 \psi^\beta_b \rangle \sim 
\Delta_1 \eps^{\alpha\beta 1}\eps_{ab1} \!+\! 
\Delta_2 \eps^{\alpha\beta 2}\eps_{ab2} \!+\! 
\Delta_3 \eps^{\alpha\beta 3}\eps_{ab3}
\label{condensate}
\end{equation} 
Here $\psi^\alpha_a$ is a quark of color $\alpha=(r,g,b)$ 
and flavor $a=(u,d,s)$;
the condensate is a Lorentz scalar, antisymmetric in Dirac indices,
antisymmetric in color 
(the channel with the strongest
attraction between quarks), and consequently
antisymmetric in flavor. The gap parameters
$\De_1$, $\De_2$ and $\De_3$ describe down-strange,
up-strange and up-down Cooper pairs, respectively. 

To find which
phases occur in realistic quark matter, one must
take into account the strange quark mass and equilibration
under the weak interaction, and impose neutrality
under the color and electromagnetic gauge symmetries.
The arguments that favor (\ref{condensate})
are unaffected by these considerations, but there is no reason for
the gap parameters to be equal once $M_s\neq 0$.
Previous work~\cite{Alford:1999pa,Alford:2002kj,Steiner:2002gx,Neumann:2002jm}
compared the 
color-flavor-locked (CFL) phase (favored in the limit $M_s\to 0$
or $\mu\to\infty$),
and the two-flavor (2SC) phase 
(favored in the limit $M_s\to \infty$).
In this paper we show that in fact a
transition between these phases does not occur.
Above a critical $M_s^2/\mu$, the CFL phase gives
way to a new ``gapless CFL phase'', not to the 2SC phase.
The relevant phases are
\begin{eqnarray}
\De_3 \!\simeq\! \De_2 \!=\! \De_1 \!=\!\De_{CFL}
  &\ \ \ & \mbox{CFL} \label{CFL} \\
\De_3>0,\quad  \De_1 \!=\! \De_2 \!=\! 0
  &\ \ \ & \mbox{2SC} \label{2SC} \\
\De_3 > \De_2 >\De_1>0  &\ \ \ & \mbox{gapless CFL} \label{gCFL}
\end{eqnarray}

To impose color neutrality, it is sufficient to
consider the $U(1)_3 \times U(1)_8$ subgroup of the 
color gauge group generated by the Cartan subalgebra $T_3 =
\diag(\half,-\half,0)$ and $T_8=\diag(\third,\third,-\twothirds)$ in
color space \cite{Alford:2002kj}.  We
introduce chemical (color-electrostatic) potentials $\mu_3$ and
$\mu_8$ coupled to the color charges $T_3$ and $T_8$, and an
electrostatic potential $\mu_e$ coupled to $Q_e$ which is the {\em negative}
of the electric charge $Q=\diag(\twothirds,-\third,-\third)$
($\mu_e>0$ corresponds to a density of electrons; $\mu_e<0$
to positrons.)
The neutrality condition on $\mu_e,\mu_3,\mu_8$ is
\beq
\frac{\p \Om}{\p \mu_e} = \frac{\p \Om}{\p \mu_3} =
\frac{\p \Om}{\p \mu_8} = 0\ .
\label{neutralityconditions}
\eeq

Any condensate of the form (\ref{condensate})
is neutral with respect to a ``rotated electromagnetism''
generated by $\tilde Q = Q -T_3 -\half T_8$, so
$U(1)_{\tilde Q}$ is never broken, but,
depending on the values of the $\De_i$,
the rest of the 
gauge group may be spontaneously broken.
We emphasize that this does not affect the neutrality condition
(\ref{neutralityconditions}): a macroscopic sample must be neutral 
under {\em all} gauge symmetries \cite{Alford:2002kj}.

Previous model-independent 
calculations~\cite{Alford:2002kj}
have compared the free energy of the CFL phase with
that of the 2SC and unpaired phases.
In the CFL phase, $\mu_e=\mu_3=0$ and $\mu_8= -M_s^2/(2\mu)$
to leading order in $M_s/\mu$. To this order, 
the CFL phase has a lower free energy than 
either 2SC or neutral unpaired quark matter for 
\beq
\frac{M_s^2}{\mu} < 4 \De_{CFL}\ .
\label{CFLcondition}
\eeq
In this paper we 
show that the CFL
phase becomes unstable already at a {\it lower} value of $M_s^2/\mu$.

We shall present
solutions to the gap equations for $\De_1$, $\De_2$ and $\De_3$ below.
First, however, 
we give a model-independent argument for
the instability of the CFL phase above some critical $M_s^2/\mu$
In the condensate~(\ref{condensate}), the $(gs,bd)$, $(bu,rs)$
and $(rd,gu)$ quarks pair with gap parameters $\De_1$, 
$\De_2$ and $\De_3$ respectively, 
while the $(ru,gd,bs)$ quarks pair among each
other 
involving all the $\De$'s.
The gap equations for the three $\De$'s are coupled, but we can,
for example, 
analyze the effect of a specified $\De_1$ on the $gs$ and $bd$ quarks 
without reference to the other quarks.
The leading effect of $M_s$ 
is like a shift in the chemical potential
of the strange quarks,
so 
the $bd$ and $gs$ quarks
feel ``effective chemical
potentials'' 
$\mu_{bd}^{\rm eff} = \mu - \twothirds \mu_8$ and
$\mu_{gs}^{\rm eff} = \mu  + \third \mu_8 -\frac{M_s^2}{2\mu}$.
In the CFL phase $\mu_8=-M_s^2/(2\mu)$~\cite{Alford:2002kj},
so $\mu_{bd}^{\rm eff} - \mu_{gs}^{\rm eff} = M_s^2/\mu$.
The CFL phase will be stable as long as the
pairing makes it energetically favorable to maintain equality of the
$bd$ and $gs$ Fermi momenta, despite their differing chemical
potentials \cite{Rajagopal:2000ff}.
It becomes unstable when
the energy gained from turning a 
$gs$ quark near the common Fermi momentum into a $bd$ quark 
(namely $M_s^2/\mu$) exceeds the cost
in lost pairing energy $2\De_1$. 
So the CFL phase is stable when
\beq
\frac{M_s^2}{\mu} < 2\De_{CFL}\ ,
\label{CFLstable}
\eeq
For larger $M_s^2/\mu$, the CFL phase is      
replaced by some new phase with unpaired $bd$ quarks,
which from (\ref{CFLcondition}) cannot be neutral unpaired
or 2SC quark matter because the 
new phase and the CFL phase must have the same 
free energy at the critical $M_s^2/\mu = 2\De_{CFL}$.

For a more detailed analysis, 
we use a NJL model 
with a pointlike four-quark interaction 
with the quantum numbers of single-gluon exchange, as
in the first paper in Ref.~\cite{reviews} but with chemical
potentials $\mu_e$, $\mu_3$ and $\mu_8$ 
introduced as in Ref.~\cite{Alford:2002kj}.
Whereas in nature, the conditions (\ref{neutralityconditions})
are enforced by the dynamics of the 
gauge fields whose zeroth components are $\mu_e$, $\mu_3$ 
and~$\mu_8$~\cite{Alford:2002kj,Gerhold:2003js},
in an NJL model Eqs.~(\ref{neutralityconditions})
must be imposed~\cite{Alford:2002kj}.
The model has two parameters, the four-fermion coupling $G$
and a three-momentum cutoff $\Lambda$, but we 
quote results in terms of the physical quantity $\De_0$ 
(the CFL gap at $M_s=0$), since varying $\Lambda$ by 20\% while tuning
$G$ to keep $\De_0$ fixed changes all of our results by at most a few
percent.
We use $\Lambda=800$~MeV in all results that we quote.

We make the ansatz (\ref{condensate}) for the diquark condensate
in the 
quark propagator, 
using the 
Nambu-Gorkov formalism, and then evaluate the 
free energy $\Omega$.  We shall present the details of our
calculation elsewhere, 
but the formalism is as in Ref.~\cite{Buballa:2001gj}, 
\new{with the additional constraints of electrical and color neutrality}.
To simplify the analysis, we neglect the color and flavor symmetric
contributions~\cite{Alford:1998mk}
\new{that respect the same symmetries and
are known to be small \cite{Alford:1998mk,Alford:1999pa}},
set the light quark masses to zero,
and treat the constituent 
strange quark mass $M_s$ as a parameter, 
as in Ref.~\cite{Alford:2002kj},
leaving a treatment like that 
of Refs.~\cite{Steiner:2002gx,Neumann:2002jm} 
in which
one solves for the $\langle \bar s s\rangle$ condensate
for the future. 
We incorporate
$M_s$ only via its leading effect, 
a shift $-M_s^2/2\mu$ in the effective chemical potential
for the strange quarks. This 
requires that $M_s^2/\mu^2$ be small,
and neglects the dependence
of the gap parameters on the Fermi velocity of the strange
quark~\cite{Kundu:2001tt}, meaning that we find
$\De_3=\De_2=\De_1$ in the CFL phase instead of finding
$\De_3$ larger than the other two by a few percent~\cite{Steiner:2002gx}.
We work to leading
nontrivial order in $\De_1$, $\De_2$, $\De_3$, $\mu_e$, $\mu_3$ 
and $\mu_8$, since these are all small compared to $\mu$.  Finally, we neglect
the effects of antiparticles.  
None of these approximations precludes a qualitative
understanding of the new phase we shall describe.

We calculate the free energy $\Omega$
and solve six coupled integral equations,
the neutrality conditions (\ref{neutralityconditions})
and the gap equations 
$\partial\Om/\partial\De_1=\partial\Om/\partial\De_2= 
\partial\Om/\partial\De_3=0$.
Our solutions depend on three parameters: $\mu$, $M_s$ and
$\De_0$.  We always take $\mu=500$~MeV, which is reasonable
for the center of a neutron star.
We quote results only for $\De_0=25$~MeV, 
which is within the plausible range~\cite{reviews}
and ensures that the transition (\ref{CFLstable}) occurs where
$M_s^2/\mu^2$ corrections are under 
control. 
Although we have obtained our results by varying $M_s$ at
at fixed $\mu$,
we typically quote results in terms of the important combination 
$M_s^2/\mu$. Note that in nature $M_s$ increases with decreasing $\mu$.

\begin{figure}[t]
\begin{center}
\includegraphics[width=0.4\textwidth]{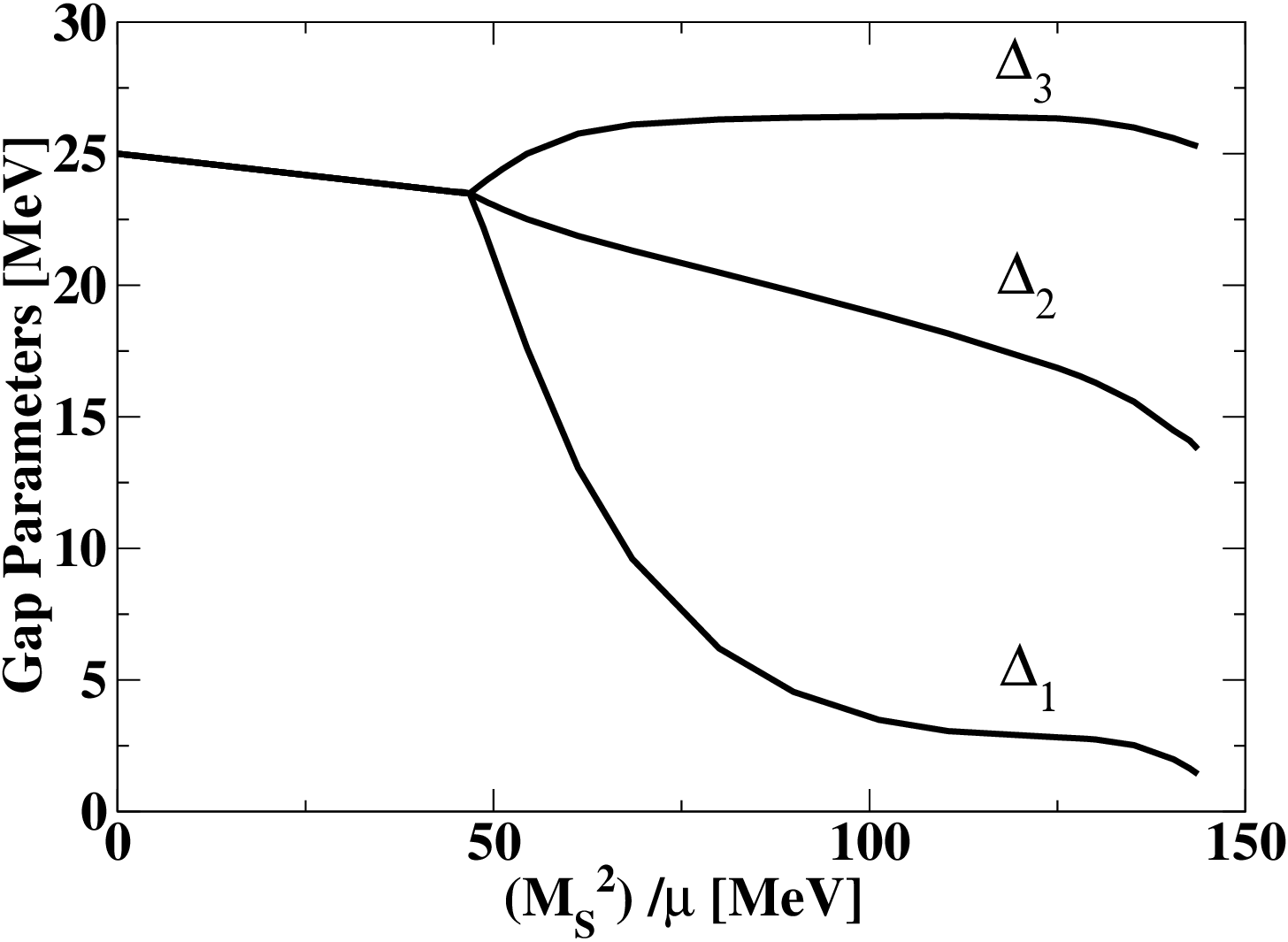}
\end{center}
\vspace{-0.25in}
\caption{
Gap parameters $\De_3$, $\De_2$, and $\De_1$
as a function of $M_s^2/\mu$ for $\mu=500$~MeV, in a model
where $\De_0=25$~MeV (see text). 
There is a \new{continuous}
transition between the CFL phase 
and the gapless CFL phase. 
}
\label{fig:gaps}
\end{figure}

In Fig.~\ref{fig:gaps}, we show the gaps as a function
of $M_s^2/\mu$, for 
$\De_0=25$~MeV. We see
a phase transition occurring at a critical
$M_s^c$ that, in our model
calculation with $\mu=500$~MeV, lies between $M_s=153$~MeV and 
$M_s=154$~MeV.  
Below $(M_s^2/\mu)_c$, 
i.e.~at high enough
density, we have the CFL phase. 
At $M_s=153$~MeV, 
$\De_1=\De_2=\De_2=23.5$~MeV and $M_s^2/\mu=46.8$~MeV:
the model-independent prediction (\ref{CFLstable}) is in 
good agreement with our
model calculation.  For $M_s^2/\mu > (M_s^2/\mu)_c$, i.e.~at densities
below those where the  CFL phase is stable, we find the
gapless CFL (gCFL) phase with $\De_3>\De_2>\De_1>0$, 
and all the gaps changing
much more rapidly with $M_s^2/\mu$.
We have checked
that upon varying $\De_0$, the critical $M_s^2/\mu$ changes
quantitatively as predicted by (\ref{CFLstable}) and
our results are otherwise qualitatively unchanged.

By evaluating the free energy, we have confirmed that the 
CFL $\to$ gCFL transition at $(M_s^2/\mu)_c$ is \new{not first order},
and found a first-order
gCFL $\to$ unpaired quark matter transition
at $M_s^2/\mu\approx 129$~MeV.
The (now metastable) gCFL phase 
continues to exist up to
$M_s^2/\mu\approx 144$~MeV where, as we show below, it ceases
to be a solution.

\begin{figure}[t]
\begin{center}
\includegraphics[width=0.4\textwidth]{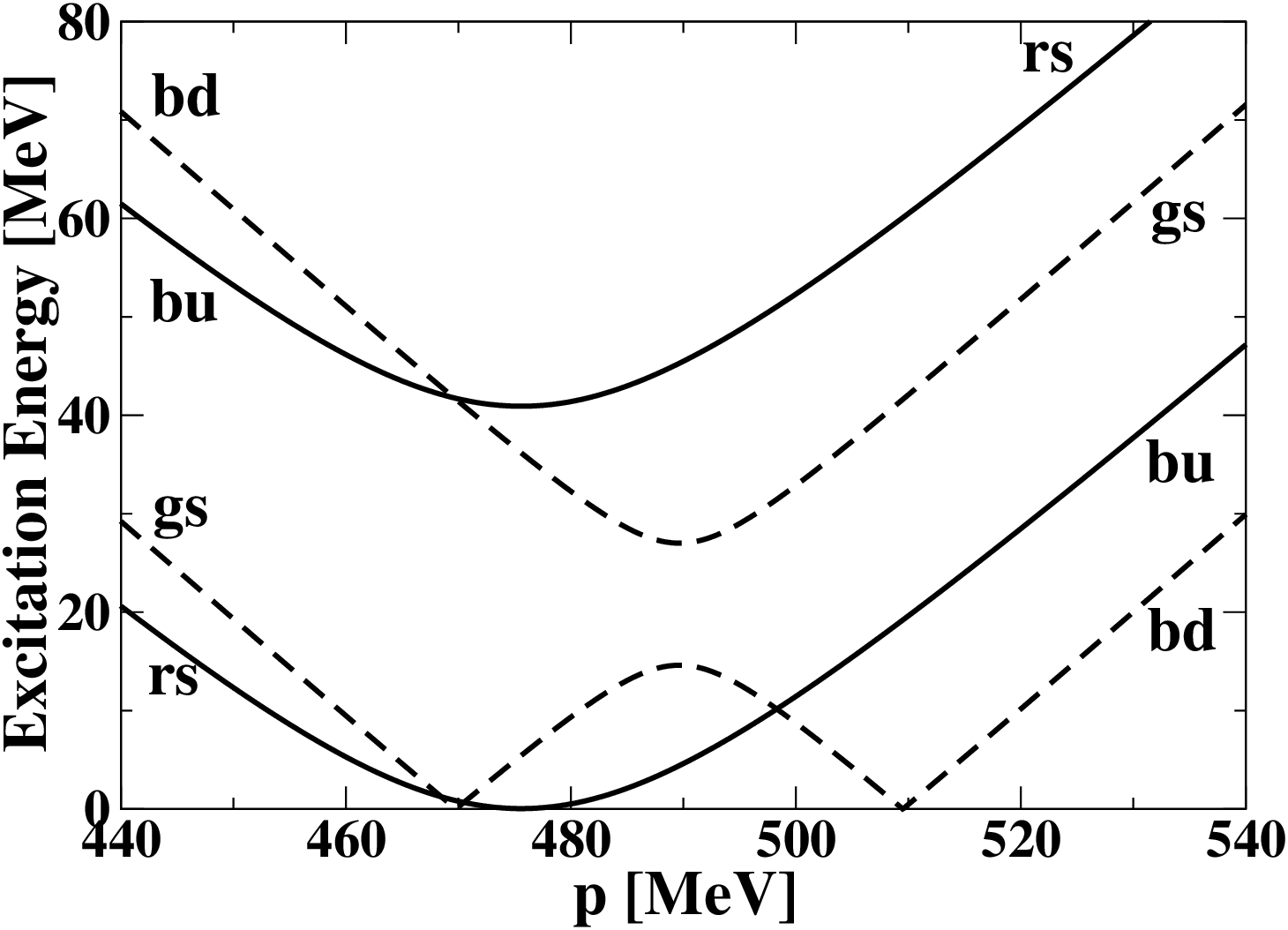}
\end{center}
\vspace{-0.25in}
\caption{Dispersion relations at $M_s^2/\mu=80$~MeV
for $gs$ and $bd$ quarks (dashed lines)
and for $bu$ and $rs$ quarks (solid lines).
There are gapless $gs$-$bd$ modes at
$p_1^{bd}=469.8$~MeV and $p_2^{bd}=509.5$~MeV, which are
the boundaries of the $bd$-filled 
``blocking''~\cite{Alford:2000ze,Bowers:2001ip} 
or 
``breached pairing''~\cite{Gubankova:2003uj} region.
One $bu$-$rs$ mode is gapless
with an almost exactly quadratic dispersion relation. The five quark 
quasiparticles not plotted all have gaps, throughout
the CFL and gCFL phases.
}
\label{fig:disprel}
\end{figure}

We see 
from the $bd$-$gs$ quasiquark dispersion relations 
(dashed lines in Fig.~\ref{fig:disprel})
that 
there are gapless excitations at momenta
$p^{bd}_1$ and $p^{bd}_2$.  The analysis of Ref.~\cite{Bowers:2001ip}
demonstrates that, as expected from 
the model-independent argument above, 
these bound a ``blocking region''~\cite{Alford:2000ze}
$p^{bd}_1<p<p^{bd}_2$ 
in which there are $bd$ quarks but no $gs$ quarks, and thus no pairing. 
We have confirmed this by explicit calculation of number densities.
The volume of the blocking region is zero at the critical point, 
and grows in the gCFL phase
like $(p^{bd}_2-p^{bd}_1)\sim [M_s^2/\mu - (M_s^2/\mu)_c]^{1/2}\De_1^{1/2}$.
Note that 
the dispersion relations in the blocking region are nontrivial
because the excitations obtained by 
either adding a $gs$ quark or removing a $bd$ quark
mix via the $\De_1$ condensate.
If we neglect this mixing,
the gapless excitations near $p^{bd}_2$ ($p^{bd}_1$)
are $bd$ ($gs$) quarks and holes.
The gapless CFL phase
is analogous to the unstable Sarma phase~\cite{Sarma},
but is rendered stable by the neutrality constraint.
This possibility, leading to a gapless phase appearing
at a \new{continuous} phase transition,
was first noted in the two flavor 
case~\cite{Shovkovy:2003uu,Gubankova:2003uj},
and it was conjectured that the three flavor case could be 
similar~\cite{Gubankova:2003uj}.
Our model also has a gapless 2SC phase, 
but it is free-energetically
disfavored relative to the gCFL phase
except perhaps 
very near where the
gCFL and unpaired quark matter free energies cross at
$M_s^2/\mu=129$~MeV: a calculation
with fewer approximations will be required to resolve a possible
small gapless 2SC ``window''.
     
\begin{figure}[t]
\begin{center}
\includegraphics[width=0.4\textwidth]{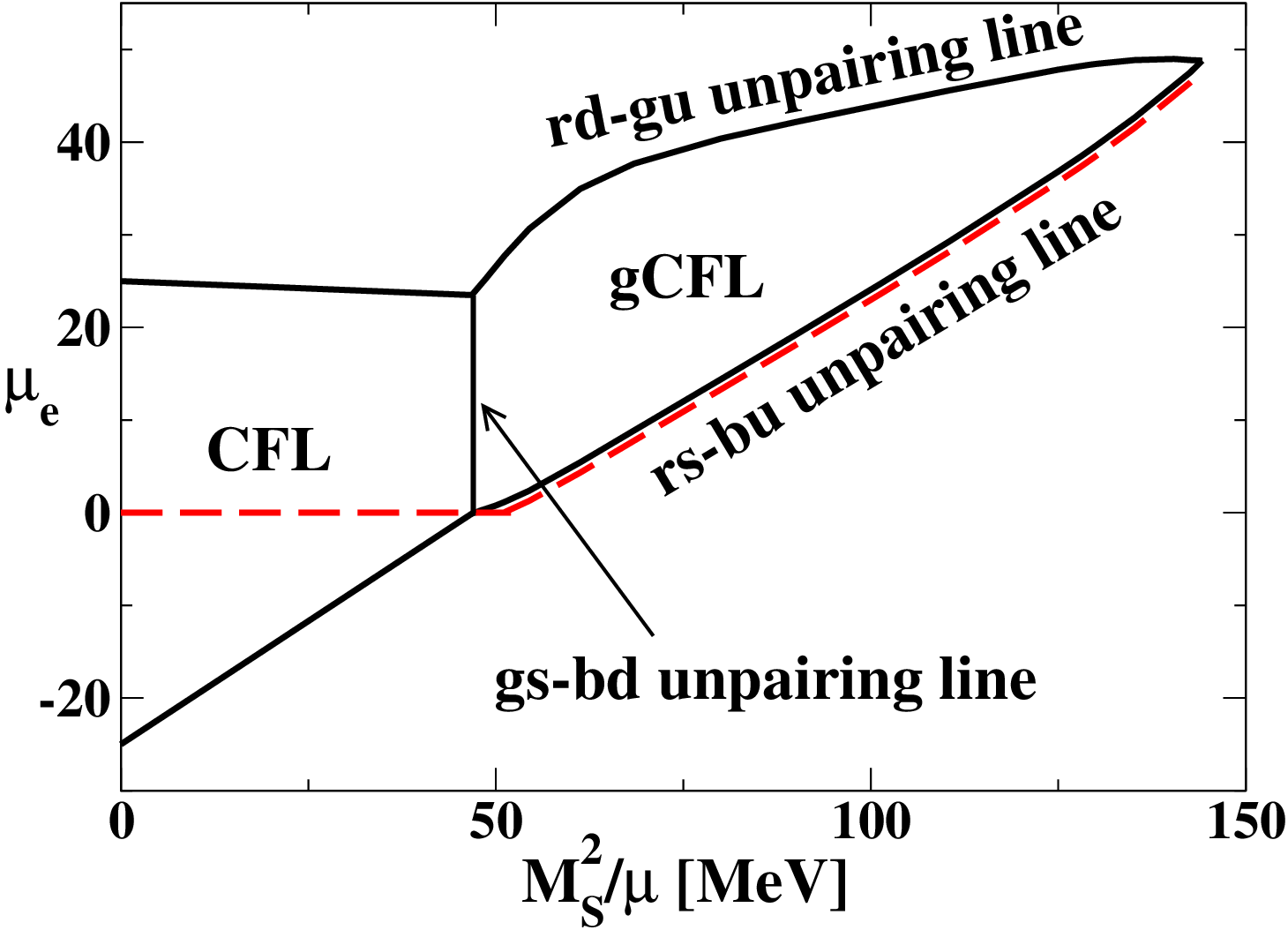}
\end{center}
\vspace{-0.25in}
\caption{The upper and lower curves bound the region of $\mu_e$ 
where CFL or gCFL solutions are found, if electrons are ignored.
Between the curves the quark matter is a $\Qtilde$-insulator.
Taking electrons into account, the correct solution
has $\mu_e=0$ for $M_s^2/\mu<(M_s^2/\mu)_c$ 
in the CFL phase (dashed line), and has  $\mu_e$  below but
{\it very} close to the lower curve for $M_s^2/\mu>(M_s^2/\mu)_c$
in the gCFL phase (see text).
}
\label{fig:mue}
\end{figure}

The $\Qtilde$-neutral $gs$ and $bd$ quarks are not the only source of
gapless modes in the gCFL phase: 
there are also gapless $\Qtilde$-charged modes, 
associated with the $bu$ and $rs$
quarks, making the gCFL phase
a $\Qtilde$-conductor rather than an insulator. Electrons
play a crucial role in this, but let us first understand the
quark matter on its own, setting the electron mass to infinity
and in so doing keeping the gCFL phase a $\tilde Q$-insulator.

In the absence of electrons, both the CFL and gCFL phases have a
degenerate set of neutral free energy minima over a range of 
$\mu_\Qtilde = -{{\txt \frac{4}{9}}}
(\mu_e + \mu_3 + \half \mu_8)$,
with the two orthogonal chemical potentials and the three
gap parameters fixed. The limits
of this range (giving $\mu_e$ rather than $\mu_\Qtilde$) are shown
in Fig.~\ref{fig:mue}.
Because $\Omega$ is independent of $\mu_{\tilde Q}$ in this range, the material
is a $\tilde Q$-insulator.  
At the upper limit in $\mu_\Qtilde$ (lower limit in $\mu_e$),
the $bu$ and $rs$ quarks, which have $\tilde Q=+1$ 
and $\tilde Q=-1$ respectively, start to unpair: they
develop a blocking region of unpaired $bu$ quarks
bounded by gapless modes,
meaning that the $\tilde Q$-neutrality condition cannot be satisfied.
At the lower limit in $\mu_\Qtilde$ (upper limit in $\mu_e$)
an analogous dielectric breakdown occurs with
the $rd-gu$ pairs breaking and a blocking
region of unpaired
$gu$ quarks with $\tilde Q=-1$ developing.
The solid curves in Fig.~\ref{fig:mue} thus define 
the ``band gap'' for $\tilde Q$-charged fermionic
excitations.
In the region between the curves, $\Qtilde$-insulating
CFL or gCFL quark matter 
exists, but outside that region no neutral solution exists.
At $M_s^2/\mu=144$~MeV, which is so large that the gCFL
phase is anyway already metastable with respect to unpaired
quark matter, the two boundaries cross, meaning that no gCFL
solution can be found.  

In the real world there are electrons, which we take to be massless.
Consequently, in the CFL phase $[M_s^2/\mu<(M_s^2/\mu)_c]$, 
neutrality requires
$\mu_e=0$, so no electrons are present and
the material remains $\tilde Q$-neutral and a $\tilde Q$-insulator, 
as before~\cite{kaoncondensation}. 
However, in the gCFL phase, $[M_s^2/\mu>(M_s^2/\mu)_c]$, $\mu_e=0$ is
below the allowed range. In this case,
the true solution lies ``just below'' the lower curve in Fig.~\ref{fig:mue},
where the $bu$ and $rs$ quarks have become gapless, allowing
a small density of unpaired $bu$ ($\tilde Q=+1$) quarks to cancel
the charge of the electrons. The density of electrons is
$\mu_e^3/(3\pi^2)$, and the density of unpaired $bu$ quarks is
$(p^{bu}_2)^3-(p^{bu}_1)^3/(3\pi^2)$, so they cancel when
$(p^{bu}_2 - p^{bu}_1) = \mu_e^3/3 \bar p^2$,
where $\bar p$ is the average of the momenta $p^{bu}_1$ and $p^{bu}_2$
that bound the $bu$ blocking region.
At $M_s^2/\mu=80$~MeV, where $\mu_e=14.6$~MeV at the lower curve
in Fig.~\ref{fig:mue}, this implies $(p^{bu}_2 - p^{bu}_1)=0.0046$~MeV!
(To resolve $p^{bu}_2-p^{bu}_1$, we solved 
the 
equations assuming 200 and 500 ``flavors'' of massless electrons.)
Because  $(p^{bu}_2 - p^{bu}_1)$ is {\it so} small, 
at the true $\tilde Q$-neutral solution
$\mu_e$ 
is {\it very} close 
to the lower curve in Fig.~\ref{fig:mue},
and the gaps are almost unaffected by the inclusion of electrons.
However, the effect of including electrons
is profound: because $\mu_e$ cannot be zero, gCFL quark matter
is deformed slightly away from being a $\tilde Q$-insulator,
so that it can carry a positive $\tilde Q$-charge to compensate
the negatively charged electrons. The gCFL phase is therefore
a $U(1)_{\tilde Q}$ conductor. The quantum phase
transition at $M_s^2/\mu=(M_s^2/\mu)_c$ is a ``metal-insulator transition'',
\new{with electron density $n_e\sim (\mu_e^2-m_e^2)^{3/2}$
as the most physically relevant order parameter.
The phase transition is continuous but higher-than-second
order ($d n_e/d(M_s^2/\mu)$ is continuous).}

The phenomenology of gCFL quark matter in compact stars
will be dominated by the modes with energy less than or of order
the temperature, which is in the range keV to hundreds of keV.
For the $gs$-$bd$ quasiparticles (Fig.~\ref{fig:disprel}),
the gapless $\tilde Q =0$ quasiparticles at $p^{bd}_1$ and $p^{bd}_2$
have conventional linear dispersion relations, except for
$M_s^2/\mu\rightarrow (M_s^2/\mu)_c$
where $p^{bd}_2-p^{bd}_1\rightarrow 0$.
For the $bu$-$rs$ quarks, the dispersion relation is strictly speaking
also quadratic only at the quantum critical point, but the gapless points
separate so slowly in the gCFL phase that this dispersion relation 
remains very close to quadratic.
For example, at $M_s^2/\mu=80$~MeV 
the maximum in the quasiparticle energy
between $p^{bu}_1$ and $p^{bu}_2$ is
$(p^{bu}_2-p^{bu}_1)^2/(8 \De_2 )= (\mu_e^6/72\bar p^4\De_2^2)\sim 0.13$~eV,
which is negligible at compact star temperatures.
The requirement of $\tilde Q$-neutrality naturally forces
the gapless $bu$-$rs$ dispersion relation to be (very close to) quadratic,
without requiring fine tuning to a critical point.

The low energy effective theory of the gCFL phase
must incorporate gapless fermions
which have number densities $\sim \mu^2 \sqrt{\De_2 T}$
(the gapless quarks with quadratic dispersion relation),
$\sim \mu^2 T$ (the gapless quarks with linear dispersion relations), 
and $\sim \mu_e^2 T$ (the electrons). In contrast, the (pseudo-)Goldstone
bosons present in both the CFL and gCFL phases have
number densities at most $\sim T^3$.
This means the gCFL phase will have very different phenomenology.
It will be particularly interesting
to compute the cooling of a compact star with a gCFL
core, because neutrino emission will be associated
with a weak $bd \leftrightarrow bu$ transition, requiring conversion
between quasiparticles with linear and quadratic 
dispersion relations.

Although we have studied
the gCFL phase in a model, all of the qualitative
features that we have focussed on appear robust, and we
have also offered a model-independent argument for the instability
that causes the transition.  It remains a possibility, however,
that the CFL gap is large enough that baryonic matter
supplants the CFL phase before $M_s^2/\mu>2\De$.
Assuming that the gCFL phase does replace the
CFL phase, it is also possible that gaps are small enough that
a third phase of quark matter could supplant the gCFL
phase at still lower density, 
before the transition to baryonic matter.
In our analysis with $\De_0=25$~MeV, this third phase would be the unpaired
quark matter at $M_s^2/\mu>129$~MeV but, unlike
our central results, this is model dependent. Other possibilities
include the gapless 2SC phase~\cite{Shovkovy:2003uu},
a three-flavor extension of
the crystalline color superconducting
phase~\cite{Kundu:2001tt,Alford:2000ze,Bowers:2001ip,Bowers:2002xr}, 
or weak pairing
between quarks with the same flavor~\cite{Alford:2002rz}.

We acknowledge helpful conversations with J. Bowers,
M. Forbes, E. Gubankova, J. Kundu,
W. V. Liu, S. Reddy, T. Sch\"afer and F. Wilczek.
Research supported in part by DOE grants 
DE-FG02-91ER40628 and DF-FC02-94ER40818.

\end{document}